\documentclass[twocolumn,prb]{revtex4}

\usepackage{graphicx}
\usepackage{epsfig}
\usepackage{dcolumn}
\usepackage{bm}
\usepackage{amsmath}

\begin{document}
\title{Current-voltage characteristics of semiconductor/ferromagnet junctions in the spin blockade regime}
\author{Yuriy V. Pershin}
\author{Massimiliano Di Ventra}
\affiliation{Department of
Physics, University of California, San Diego, La Jolla, California
92093-0319}

\begin{abstract}
It was recently predicted [Phys. Rev. B {\bf 75}, 193301 (2007)]
that spin blockade may develop at nonmagnetic semiconductor/perfect
ferromagnet junctions when the electron flow is directed from the
semiconductor into the ferromagnet. Here we consider current-voltage
characteristics of such junctions. By taking into account the
contact resistance, we demonstrate a current stabilization effect:
by increasing the applied voltage the current density through the
junction saturates at a specific value. The transient behavior of
the current density is also investigated.
\end{abstract}

\maketitle

There is currently a great deal of interest in spin-dependent
transport phenomena in semiconductors and their junctions with
ferromagnets~\cite{review,per,Flatte,saikin,bleibaum,rashba,pershin,r5,r6,r7,r8,r8a,stephens,crooker,sham1}.
In large part, this interest is motivated by the goal to exploit
these phenomena in new technologies, such as
spintronics and quantum computation~\cite{review}.
Recently, some attention has been focused on the problem of the {\em
extraction} of spin-polarized electrons from the semiconductor to
the ferromagnet~\cite{r8,r8a,stephens,crooker,sham1}. Despite
the apparent similarity of spin extraction with spin injection, spin
extraction shows unique features. In particular, we have recently
predicted~\cite{per} that the spin extraction process at nonmagnetic
semiconductor/perfect ferromagnet junctions can be limited by spin
blockade. The physical mechanism for spin blockade is the following:
the outflow of majority-spin electrons from the semiconductor leaves
a cloud of minority-spin electrons, which limits the majority-spin
current through the junction.

In this letter we explore consequences of this phenomenon that can
be easily verified experimentally. In particular, we study the
current-voltage characteristics of such junctions. We do so by
considering the conductivity of each of its components
(semiconductor, ferromagnet and their contact). We show that the
current flowing in a circuit involving a semiconductor/ferromagnet
interface in the spin blockade regime saturates with increasing
applied voltage. Therefore, such an interface can be potentially
used as a {\it spin-based current stabilizer}. We also show that
in structures with a semiconducting region longer that the spin
diffusion length, the current density saturates to the critical
current density $j_c$ found in Ref.~\onlinecite{per}. Instead, in
junctions with the semiconductor region shorter than the spin
diffusion length, the asymptotic current value mey be different
from $j_c$ depending on how the semiconductor is connected from
the opposite side of the junction. In particular, if this second contact is a good
contact with a normal metal, then the asymptotic current value is
higher than $j_c$. We also consider transient processes, which,
due to the finite response time of the spin polarization to the
applied voltage, limit the speed of operation of such devices.

\begin{figure}[b]
\includegraphics[angle=0,width=8cm]{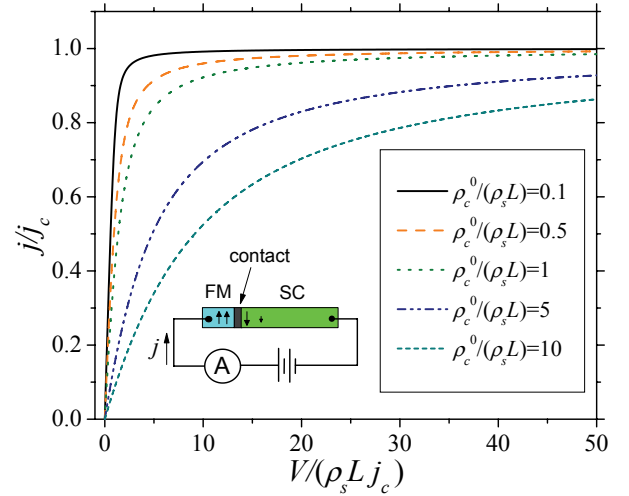}
\caption{\label{fig1} (Color online) Current-voltage characteristic
of the system calculated for several values of the ratio between the
contact resistance and the semiconductor resistance
$\rho_c^0/(\rho_sL)$. $j_c=eN_0\sqrt{D/(2\tau_{sf})}$ is the
critical current density, $-e$ is the electron charge, $N_0$ is the
electron density in the semiconductor, $D$ is the diffusion
coefficient, and $\tau_{sf}$ is the spin relaxation time. Other
symbols are defined in the text. Inset: schematic drawing of the
circuit.}
\end{figure}

The circuit we have in mind is shown schematically in the inset of
Fig. \ref{fig1}. We consider a voltage source (battery) connected to
the semiconductor and ferromagnet regions of the junction. Assuming
that the ferromagnet is a good conductor we can neglect the voltage
drop across it. We also assume a good contact of the voltage source
with the semiconductor (ohmic or nonlinear contact at this junction
can be easily incorporated into our model). Therefore, there are two
components of the total circuit where the voltage mainly drops: the
semiconductor part, and its contact with the ferromagnet. We can
then write the total applied voltage $V$ as $V=V_s+V_c$, where $V_s$
and $V_c$ are voltage drops across the semiconductor region, and the
contact, respectively. In our model, we consider a perfect
ferromagnet, such as a half-metal ferromagnet. While both spin-up
and spin-down electrons are injected from the battery into the
semiconductor, only, let say, spin-up electrons are extracted from
the semiconductor into the ferromagnet.

Spin and charge transport of a non-degenerate electron gas in the
semiconductor can be conveniently described within the
drift-diffusion approximation~\cite{Flatte,per}. For simplicity, we
neglect charge accumulation effects as in Ref. \onlinecite{per}. In
the semiconductor region we can then write

\begin{equation}
j=\sigma E=eN_0\mu\frac{V_s}{L}\equiv \frac{V_s}{\rho_s L},
\label{js}
\end{equation}
where $j$ is the current density, $\sigma$ is the conductivity, $E$
is the electric field, $\mu$ is the mobility defined via $\vec
v_{drift}=\mu \vec E$, $L$ is the length of the semiconductor, and
$\rho_s$ is the semiconductor resistivity. Next, we consider the
voltage drop across the contact. The conductivity of the contact is
proportional to the density of majority spin electrons in the
semiconductor near the contact, $n_\uparrow (0)$. Therefore,
assuming a linear relationship between the current and voltage drop
across the contact at a fixed spin-up density, $n_\uparrow (0)$, we
write

\begin{figure}[tb]
\includegraphics[angle=0,width=8cm]{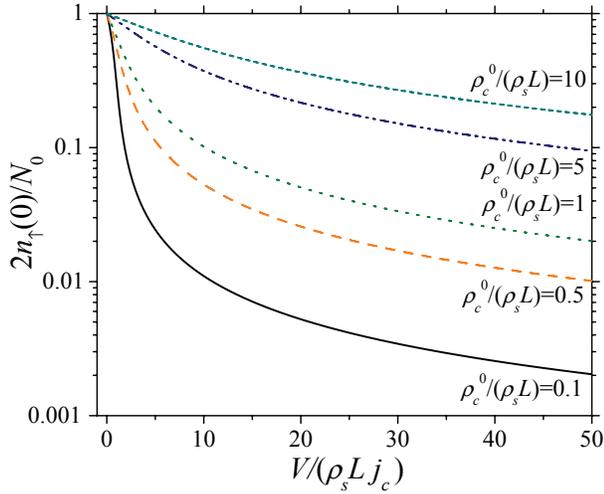}
\caption{\label{fig2} (Color online) Spin-up density at the
junction as a function of the applied voltage for several values
of $\rho_c^0/(\rho_sL)$.}
\end{figure}

\begin{equation}
j=\frac{2 n_\uparrow (0)}{N_0}\frac{V_c}{\rho_c^0},  \label{jj}
\end{equation}
where $\rho_c^0$ is the steady-state contact resistivity at
$V\rightarrow 0$ (when $n_\uparrow(0)=N_0/2$). Combining Eqs.
(\ref{js},\ref{jj}) we get

\begin{equation}
V=V_s+V_c=\left(\rho_s L+\rho_c^0\frac{N_0}{2 n_\uparrow (0)}
\right)j.\label{IVeq}
\end{equation}
Eq. (\ref{IVeq}), which couples $V$ and $j$, must be supplemented by
the system of drift-diffusion equations for the semiconductor region
whose solution gives $n_\uparrow (0)$. This system of equations
consists of the continuity equations for spin-up and spin-down
electrons, and the equations for the two spin currents:

\begin{equation}
e\frac{\partial n_{\uparrow (\downarrow)}}{\partial
t}=\textnormal{div} \vec j_{\uparrow
(\downarrow)}+\frac{e}{2\tau_{sf}}\left(n_{\downarrow
(\uparrow)}-n_{\uparrow (\downarrow)} \right), \label{contEq}
\end{equation}
\begin{equation}
\vec j_{\uparrow (\downarrow)}=\sigma\vec E+eD\nabla n_{\uparrow
(\downarrow)}. \label{currentEq}
\end{equation}
It is assumed that the total electron density in the semiconductor
is constant, i.e., $n_{\uparrow}(x)+n_{\downarrow}(x)=N_0$.
Correspondingly, the electric field is homogeneous and coupled to
the total current density as $j=e\mu N_0E_0$. The boundary
conditions are: $j_\uparrow (0)=j$, $j_\downarrow (0)=0$,
$n_\uparrow (L)=n_\downarrow (L)=N_0/2$.

In the following, we will consider separately the two cases of long
($L\gg l_s$) and short ($L\lesssim l_s$) semiconductor regions, with
$l_s$ the spin diffusion length defined below.

\begin{figure}[b]
\includegraphics[angle=0,width=8cm]{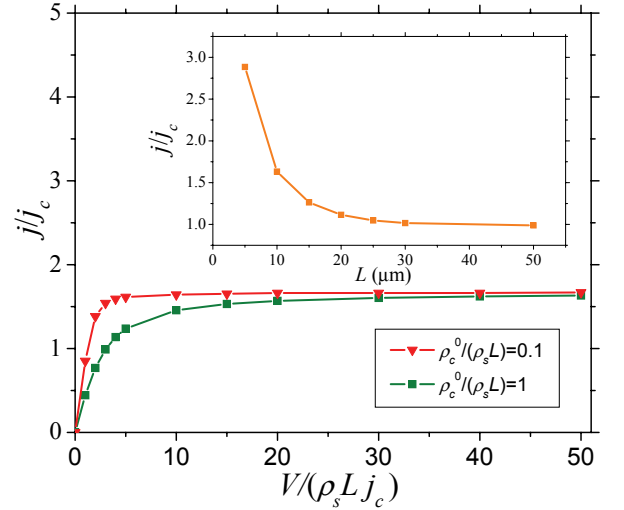}
\caption{\label{fig3} (Color online) Current-voltage
characteristic of a junction with $L=10\mu$m showing the current
density saturation at $j>j_c$. Inset: steady-state current density
as a function of $L$ at the fixed value of $V/(\rho_sLj_c)=50$.
These plots were obtained using parameter values $D=220$cm$^2$/s,
$\mu=8500$cm$^2$/(Vs), $N_0=5\cdot 10^{15}$cm$^{-3}$ and
$\tau_{sf}=10$ns.}
\end{figure}

{\it (i) $L\gg l_s$}.--- In this limit, a steady-state solution of
Eqs. (\ref{contEq},\ref{currentEq}) is known~\cite{per}. The spin
densities decay exponentially from the junction to their bulk values
of $N_0/2$. The decay occurs on the length scale of the up-stream
spin-diffusion length~\cite{Flatte,per} $l_s=2D/\left( \mu E_0
+\sqrt{\mu^2 E_0^2+4D/\tau_{sf}} \right)$. The spin-up density at
the junction is~\cite{per}

\begin{equation}
n_\uparrow(0)=\frac{N_0}{2}-\frac{N_0}{\sqrt{1+4\frac{D}{\tau_{sf}\mu^2E_0^2}}-1}.
\label{spinup}
\end{equation}
Substituting Eq. (\ref{spinup}) into Eq. (\ref{IVeq}) and
introducing the dimensionless current density $\tilde j=j/j_c$, we
get a closed equation coupling current density and voltage:

\begin{equation}
\frac{V}{\rho_s L
j_c}=\left(1+\frac{\rho_c^0}{\rho_sL}\frac{1}{1-\frac{2}{\sqrt{1+\frac{8}{\tilde
j^2 }}-1}} \right) \tilde j. \label{IVclosed}
\end{equation}

Fig. \ref{fig1} shows solutions of Eq. (\ref{IVclosed}) at
different values of the ratio of the contact resistance to the
resistance of the semiconductor region. All curves saturate at
$j/j_c=1$ with increasing voltage. The saturation occurs faster in
systems having smaller contact resistance. In Fig. \ref{fig2}, we
plot the corresponding spin-up density $n_\uparrow (0)$. It
follows from Figs. \ref{fig1} and \ref{fig2} that, for the
selected values of parameters, the current density $j$ is quite
close to the critical current density $j_c$ at voltages for which
$2n_\uparrow (0)/N_0\sim 10^{-2}$. For current stabilization
applications, by specifying the maximum desired deviation of $j$
from $j_c$, one can obtain the minimal voltage $V_{min}$ required
for that deviation using Eq. (\ref{IVclosed})~\cite{prec}.

{\it (ii) $L\lesssim l_s$.}--- In this limit, Eqs.
(\ref{contEq},\ref{currentEq}), supplemented by Eq. (\ref{IVeq}),
were solved numerically~\cite{numscheme}. Starting with
unpolarized electrons in the semiconductor, we have iterated at
each time step Eqs. (\ref{contEq},\ref{currentEq}) with the
constrain imposed by Eq. (\ref{IVeq}).

In this regime, the current-voltage characteristics have a similar
saturation behavior as in the case $L\gg l_s$. However, the
asymptotic values of the current density ($t\rightarrow \infty$,
$V\rightarrow \infty$) are higher than $j_c$ (see Fig. \ref{fig3}).
This is due to the boundary conditions $n_\uparrow (L)=n_\downarrow
(L)=N_0/2$. Such boundary condition describes a perfect contact of
the semiconductor with a large reservoir of spin-unpolarized
electrons. These spin-unpolarized electrons facilitate diffusion of
electrons from the contact region, reducing the level of spin
polarization near the contact and thus increasing the current
density at which spin blockade occurs. We plot the current density
as a function of $L$ in the inset of Fig. \ref{fig3}. For the
selected set of parameters, the current density starts to deviate
noticeably from $j_c$ in structures with $L\lesssim 20\mu$m.

\begin{figure}[t]
\includegraphics[angle=0,width=8cm]{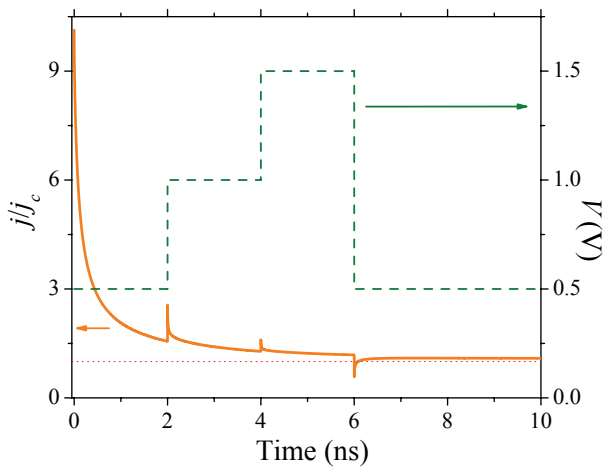}
\caption{\label{fig4} (Color online) Transient current (solid
line) excited by application of step voltages (dashed line). The
dotted horizontal line corresponding to $j=j_c$ is a guide for
eye. Here we used $L=20\mu$m and $\rho_c^0/(\rho_sL)=1$. The rest
of parameters are as in Fig. \ref{fig3}.}
\end{figure}

Finally, in view of potential applications, it is important to
know the transient behavior of the current density. To do this, we
consider stepwise voltage changes as shown in Fig. \ref{fig4}.
This illustrative shape of $V$ was selected to show the response
to both positive and negative voltage increments. The resultant
current density depicted in Fig. \ref{fig4} exhibits spikes at
each change in $V$. The main change in current density occurs
during the first several hundreds of picoseconds after the voltage
is applied. Physically, during this time period the electron spin
polarization adjusts to a new value of the bias. In particular,
immediately after an increase of $V$, $n_\uparrow (0)$ is larger
than its steady-state value at the same voltage. Therefore,
accordingly to Eq. (\ref{IVeq}), a positive spike in $j$ appears.
Similarly, a stepwise decrease of $V$ results in a negative spike.
We finally note that current density spikes can not be fitted by a
single exponent.

This work is partly supported by the NSF Grant No. DMR-0133075.


\begin{thebibliography}{}
\bibitem{prec} We note that $V_{min}$ is linear in $\rho_c$.
\end{thebibliography}
\end{document}